# Power and Delay Aware On-Demand Routing For Ad Hoc Networks

Alok Kumar Jagadev
SOA University, Jagamohan Nagar,
Jagamara, Bhubaneswar – 751030
E-mail address : a_jagadev@yahoo.co.in

Binod Kumar Pattanayak
SOA University, Jagamohan Nagar,
Jagamara, Bhubaneswar – 751030
E-mail address : bkp_iter@yahoo.co.in

Manoj Kumar Mishra
SOA University, Jagamohan Nagar,
Jagamara, Bhubaneswar – 751030
E-mail address : mkmishra_iter@yahoo.com

Manojranjan Nayak
SOA University, Jagamohan Nagar,
Jagamara, Bhubaneswar – 751030
E-mail address : president@soauniversity.ac.in

*Abstract* - **Wide implementation of IEEE 802.11 based networks could lead to deployment of localized wireless data communication environments with a limited number of mobile hosts, called ad hoc networks. Implementation of a proper routing methodology in ad hoc networks makes it efficient in terms of performance. A wide spectrum of routing protocols has been contributed by several researchers. Real time applications have been most popular among the applications, run by ad hoc networks. Such applications strictly adhere to the Quality of Service (QoS) requirements such as overall throughput, end-to-end delay and power level. Support of QoS requirements becomes more challenging due to dynamic nature of MANETs, where mobility of nodes results in frequent change in topology. QoS aware routing protocols can serve to the QoS support, which concentrate on determining a path between source and destination with the QoS requirements of the flow being satisfied. We propose a protocol, called Power and Delay aware Temporally Ordered Routing Algorithm (PDTORA), based on Temporally Ordered Routing Algorithm (TORA) Protocol, where verification of power and delay requirements is carried out with a query packet at each node along the path between source and destination. Simulations justify better performance of the proposed new protocol in terms of network lifetime, end-to-end delay and packet delivery ratio as compared to TORA.**

*Keywords -  Power, delay, on-demand routing, ad hoc networks.*

## I. INTRODUCTION

Wide implementation of IEEE 802.11 based wireless networks could lead to deployment of localized wireless data communication environments called ad hoc networks. Such networks do not support wired communication and fixed infrastructure as well. The wireless nodes in MANETs are allowed to run applications, which share data of different types and characteristics. Applications running on MANETs may possess different characteristics like network size, frequency of topology change, communication requirements and data characteristics. Every node lies within the coverage area of the MANET and can communicate with any other node in the network within its own transmission range. However, nodes are free to move within the coverage area of the MANET. A node is allowed to communicate with another node not lying within its transmission range, via multi-hop routes, where each node along the route acts as a router of the message. At the same time, new nodes can join the network any time and existing nodes can leave the network any time too.

Design of communications and routing protocols becomes a challenging factor due to dynamic nature of MANETs. One of the major challenges arises around design of multi-hop routing communication protocols. Most of the existing routing protocols such as Dynamic Source Routing (DSR) protocol, Ad hoc On-demand Distance Vector (AODV) protocol, Temporally ordered Routing Algorithm (TORA) protocol and many other such protocols mostly rely on best effort service [1]. However, a best effort service may not be able to fulfill the purposes in routing for multimedia and real time applications, which strictly require the network to adequately provide the guarantees to QoS.

A variety of routing protocols has been proposed by different authors that effectively support multi-hop communications in MANETs. Such protocols can be globally categorized as: on-demand or reactive protocols like DSR, AODV and TORA; table-driven or proactive protocols such as Destination Sequenced Distance Vector protocol (DSDV). In on-demand routing protocols, a route is established between the required source and destination prior to the communication and removed after the communication is over. In table-driven routing protocols, each node implements a routing table, which permanently stores the routing information to all possible destinations, irrespective of whether a communication is initiated or not.  In table-driven approach, latency involved in route acquisition is negligibly small. However, it includes regular updates of routing information for routes, which might not be used for a longtime, and subsequently incurs a convincing overhead. In addition, this approach needs more memory space for the routing table, as more and more routing information are appended to the routing table.

Another routing approach, called Zone Routing Protocol (ZRP), has been proposed, which incorporates the benefits of on-demand as well as table-driven approaches. It implements a proactive table-driven strategy for route establishment among the nodes of the same zone, and an on-demand reactive strategy is used for establishment of communication between nodes belonging to different zones. Such a protocol can be effectively implemented in larger ad hoc networks, where the applications exhibit a high degree of locality of communications, i.e. node with close proximity to each other communicate more frequently than the nodes lying farther.

The following sections of the paper are organized as follows. Section II covers a brief review of existing QoS aware routing protocols. In Section III, a power and delay aware TORA (PDTORA) protocol is described. Section IV comprises the simulations, Section V concludes the paper, and Section VI includes the probable future enhancements.

## II. BACKGROUND

In QoS-aware routing protocols, the principal goal is to determine a path between a source and the desired destination with the specified QoS requirements being satisfied. The basic constraints around determining a QoS-aware path are minimal search, distance and trace conditions. Obviously, a QoS-aware routing protocol is so called since it performs path selection on the basis of a specified QoS. A brief overview of QoS aware routing protocols for MANETs is included in this section.

In [2], the authors have proposed Power Aware Multiple Access (PAMAS) protocol, where a node can switch off its radio link for a specific duration of time, if it perceives that it would not be able to send or receive packets due to multiple access interferences. Authors in [3] have introduced power-aware metrics resulting in power-efficient routes. Such metrics include maximizing the time of network partition and reducing the variance in power levels of nodes. These metrics can be directly implemented in a network with a centralized control, which can possibly use a routing algorithm, based on minimizing the power level (power per bit) to transmit a packet between the source and destination. One such routing algorithm proposed by authors in [4], conditional max-min battery capacity routing algorithm, chooses a route with minimal transmission power, where all nodes along the route, possess remaining battery capacity higher than a predefined threshold. In case at least one of the nodes along a route does not satisfy to the required minimum battery capacity, the route is rejected. QoS routing protocols such as Core Extraction Distributed Ad hoc Routing (CEDAR) protocol [5] are implemented for small to medium size ad hoc networks comprising tens to hundreds of nodes, where first of all, the core of the network is dynamically established and then the link states of stable high-bandwidth links are propagated to the nodes of the core. An on-demand route computation is performed using the local state by the nodes of the core. An on-demand route computation is performed using the local state by the nodes of the core. A Time Division Multiple Access (TDMA) based computation of available bandwidth for ad hoc networks is proposed by authors of [6]. It performs end-to-end bandwidth allocation following computation. This approach enables the source node to determine the availability of resources to support the desired QoS requirements.

Authors in [7] have proposed QoS-TORA protocol, based on link reversal best effort protocol TORA, which is designed for a TDMA network. In this approach, measurement of bandwidth of a link is made in terms of slot reservations during the data phase of a TDMA frame. It is implemented in Medium Access Control (MAC) as well as network layers. Simulations demonstrate its capability to establish a route with end-to-end QoS being maintained. It would be observed from the simulations that QoS-TORA provides a better throughput under highly mobile environments. Optimized Link State Routing (OLSR) protocol based solution proposed by authors in [8], performs delay and throughput aware QoS routing, and demonstrates better results in packet delivery ratio, packet loss ratio and delay, as compared to that in OLSR.

INORA, a QoS routing protocol [9], incorporates the features of INSIGNIA and TORA. In particular, it makes use of in-band signaling mechanism of INSIGNIA and QoS routing mechanism of TORA. QoS signaling in INORA is used to reserve and release resources, to set up, tear down and renegotiate flows in the network. This signaling mechanism operates independent of TORA routing protocol. In this approach, first of all, TORA determines a route between source and destination, and then, the signaling mechanism (INSIGNIA) performs reservation of resources along the route provided by TORA.

## III. POWER AND DELAY AWARE TORA (PDTORA)

Implementation of QoS routing protocols in ad hoc networks serves to fulfill the purpose of reservation of sufficient resources along a route so as to meet the QoS requirements of a flow. On the other hand, the QoS routing protocol should be able to find the path that consumes minimum resources [10]. QoS metrics vary from application to application. Major QoS metrics for ad hoc networks are available bandwidth, cost, end-to-end delay, power, packet loss ratio and so on. The QoS metrics can be generally classified as, additive metrics, concave metrics and multiplicative metrics.

For a given link (s,d), let q(s,d) be the performance metric, with s as the source and d as the destination nodes. The path $(s, s_1, s_2, \ldots, s_k, d)$ connects s and d. A given constraint is said to be additive, if $q(s,d) = q(s,s_1) + q(s_1,s_2) + \ldots + q(s_k,d)$. Thus, end-to-end delay dl(s,d) along (s,d), is an additive constraint, since it comprises the delay incurred at each link along the path (s,d). Further, a constraint is said to be concave if $q(s,d) = \min \{ q(s,s_1), q(s_1,s_2), ....., q(s_k,d)\}$. Thus, the bandwidth requirement $bw(s,d)$, between s and d is concave, since it comprises the minimum bandwidth between the links along the path. Similarly, a constraint is multiplicative if $q(s,d) = q(s,s_1) \times q(s_1,s_2) \times ...... q(s_k,d)$. For example, the probability $P(s,d)$ of a packet, sent from s to d is multiplicative, as it is the product of probabilities of individual links along the path. Hence, bandwidth and power are concave metrics, whereas cost, delay and jitter are additive metrics. Example of a multiplicative metric [11] can be reliability on availability of a link based on certain criteria such as link breakage probability.

*3.1. Delay*

Communication delay of a packet across an ad hoc network is the latency consumed by a packet to reach the destination from the source. The components of end-to-end latency of a packet at the network layer are processing delay, packetization delay, transmission delay, queuing delay and the propagation delay. Subsequently, the end-to-end delay of a path represents the sum of delay incurred at each link along the path. Node delay involves the protocol processing time at node i for link ( i, j), and link delay is the latency consumed by the packet to travel from node i to node j, i.e. along link (i, j). For wireless ad hoc networks, propagation delays are negligibly small and almost equal for each hop along the path. The major factors involved in computation of node delay are the queuing delay and delay incurred at the MAC layer processing. Computation of MAC layer delay is elaborated in [12], and two dimension finite-state Markov models [13] can be used for estimation of queuing delay, which is determined from the queuing delay distribution $Pr(D>t)$, where the average queuing delay is defined to be D, for which delay distribution is more than 90%. Conclusively end-to-end delay of a path can be obtained by adding up the node delays and link delays along the path.

*3.2. Power*

In the route discovery phase in the on-demand routing protocols like DSR, a shortest possible path is chosen and maintained until the path breaks. Hence, usage of such a path for communication for longer period of time may result in reduction of power at the nodes along the path. It is more likely, when a node belongs to multiple active routes. It results as a consequence of transmission and reception of each message causing the battery power being drained out. When a node runs out of battery power, it is unable to forward any message along the path of communication, and consequently falls out of the network. In such a case, the route breaks, and the protocol initiates another route discovery phase to find another alternative route. Such scenarios of dying nodes may adversely affect the operational life time of the ad hoc network. The principal goal of this protocol is to perform routing around nodes with higher battery power, which enhances life time of the network. The maximum power provided by the battery of a node, when fully charged, is considered to be the initial power of the node, which is taken to be the power metric.

*3.3. Power and Delay Extension in TORA*

As a source-initiated on-demand routing protocol, TORA relies on a link reversal algorithm and provides loop-free multipath routes to a specified destination [1]. In this approach, a node maintains the topology information involving its one-hop neighbors. During a reconfiguration process following a path break, TORA has the unique property to limit the control packets to a small region. The metrics such as delay, power and distance used in TORA, are depicted in Fig. 1. For a given node n, H(n) denotes its height from the destination node. Three major functions performed by TORA are: establishing, maintaining and erasing routes. Route establishment function is initiated, when a source node requires a path to a specific destination, to which it does not possess a directed link. During this process, a destination-oriented Directed Acyclic Graph (DAG) is established using a query / update mechanism.

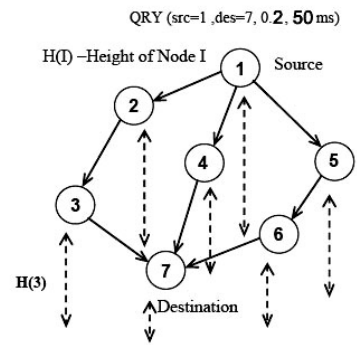

Figure 1. Power and Delay Extension in TORA

Prior to a communication, a source node sends a query packet to the destination, which incorporates the information regarding source address, destination address, minimum power level, maximum permissible delay (QRY (<source address>, <destination address>, <minimum power level>, <maximum delay>). The power extension in the query packet indicates the minimum power required to be available along the path during the communication. In addition, the delay extension specifies the maximum delay allowed between the source and destination. As depicted in Fig. 1, QoS power extension 0.2 indicates that a minimum of 20% initial power level be available along the path and a maximum allowable delay of 50 milli seconds (ms). The Verification for specified QoS power and QoS delay is made at each node as the query packet traverses the path from source to destination. A query packet is dropped if one of the constraints is not satisfied at any point of time.

As the query packet traverses the network, each node compares its available power level with the power level,

mentioned in the query packet. If the available power level at a node is found to be less than the power level specified in the query packet, then the query packet is dropped. In case the QoS power holds perfect, then the delay to destination is estimated, and if the estimate exceeds the QoS delay as mentioned in the query packet, then the packet is dropped. If the delay constraint is satisfied, the node subtracts its Node Traverse Time (NTT) from the delay bound provided in the extension and the query packet is forwarded to next hop along the route.

packet to the destination node 7, resulting in the DAG, depicted in Fig. 1. After a path to the destination is established, it is presumed to exist as long as it is required, in spite of the changes in path lengths as a result of reconfigurations, taking place during the data transfer.

In case the route to the destination is found by an intermediate node to be invalid as shown in Fig. 3, it alters its distance value to a higher value than its neighbors and originates an update packet (node 6 in Fig. 3).

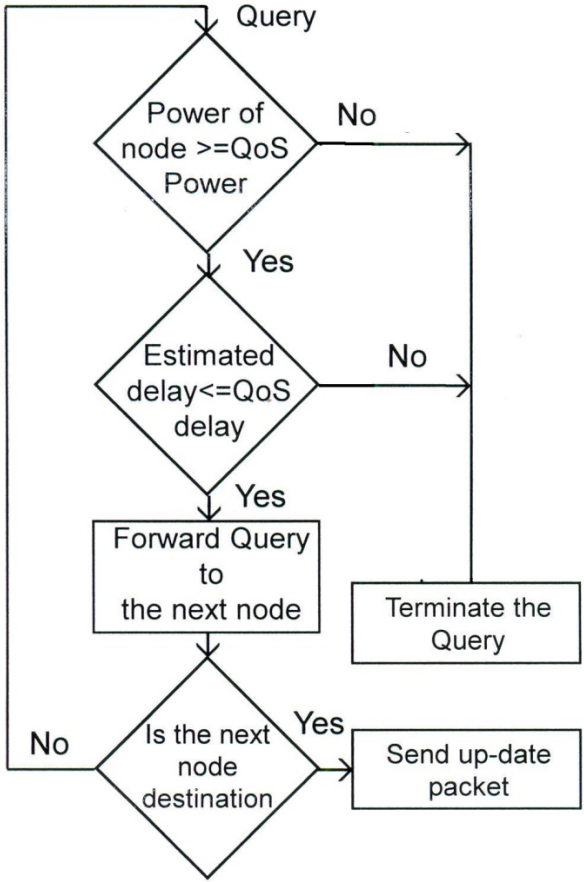

Figure 2. Algorithm of Power and Delay Extension in TORA

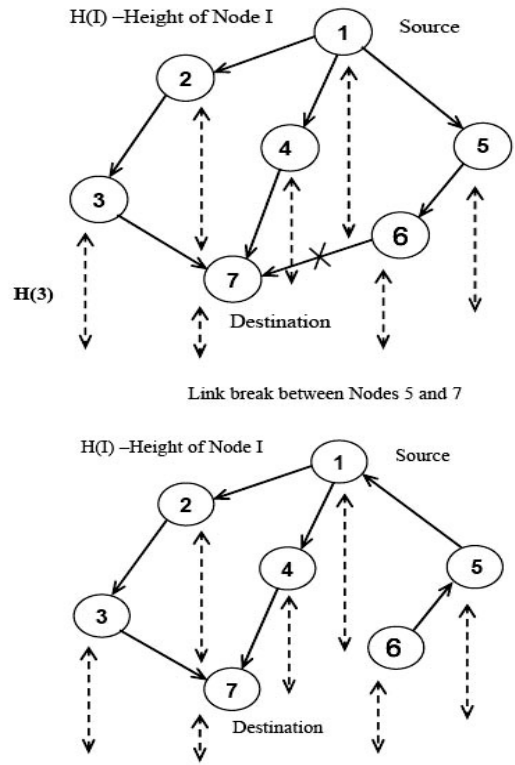

Figure 3. Route Maintenance in TORA

Fig. 2 demonstrates the sequence of operations during traversal of a query packet, which is forwarded by nodes 2,3,4,5,6 between node 1 (source) and node 7 (destination). Each node that terminates the query packet, replies with an update packet back to the source, indicating its distance from the destination and delay.

As shown in Fig. 1, the destination 7 originates an update packet. Each node along the path of this packet sets its distance to a higher value than the distance of sender of the update packet. In addition, each intermediate node adds its own NTT to the delay field of the packet. As a result, a set of directed links are created from the originator of the query

On receiving the update packet from node 6, node 5 reverses its link with node 1 and forwards the update packet to it. It results in a change in DAG as compared to Fig. 1. In case none of the neighbors of the source node has a path to the destination, it needs to initiate a fresh query / update procedure. If the link between nodes 1 & 6 breaks, then node 5 reverses its path to node 6, which is in conflict with the earlier reversal, and hence a partition in the network can be inferred (Fig. 3). When a node detects a partition, it originates a clear message, which erases the information regarding existing path in the partition to the specified destination.

## IV. SIMULATION AND PERFORMANCE EVALUATION

The proposed scheme is evaluated using ns-2 simulator [14]. It uses the random way point model for ad hoc networks. In this simulation, the ns-2 WaveLAN implementation for MAC 802.11 is used, with a channel access rate of 2 Mbps in an ad hoc network with 50 mobile nodes. Each mobile node has a mobility range of 670m x 670m. Radio transmission range of each node is set to 250m. The QoS constraints are set as 250ms for delay and 20% of initial power, with the initial power for each node being set to 20 joules, which means a combined network initial power is set to 1000 joules.

The performance metrics are chosen as follows:

Packet delivery ratio: It represents the ratio of number of packets received by the destination to the number of packets sent by the source.

Average end-to-end delay: It is defined as the end-to-end delay experienced by packets from source to destination, which includes route discovery latency, queuing delay at node, transmission delay at the MAC layer and the propagation delay across the wireless channel.

### 4.1. Packet Delivery Ratio

Packet delivery ratio for TORA and PDTORA protocols is depicted in Fig. 4, where speed of mobility taken into account is up to 100 meters/second with a pause time of 10 seconds. At low speeds of nodes, both the protocols demonstrate higher throughput. However, higher speeds may lead to frequent changes in links and probable link failures, ultimately reducing throughput. It can be observed from Fig. 4, that packet delivery ratio in PDTORA is 3% higher than that in TORA for high mobility up to 100 m/s.

Packet delivery ratio with respect to number of nodes for different mobile speeds is depicted in Fig. 5. In Fig. 5 (a), for mobile speed of 10 m/s, PDTORA shows 15% improvement over TORA. Accordingly, in Fig. 5 (b), for mobile speed of 20 m/s, PDTORA possesses 23% improvement in packet delivery ratio over TORA. Please note that in the simulation, number of nodes is set up to 50.

### 4.2. End-to-end delay

A measure of end-to-end delay for the QoS requirement of 250 ms with different node mobility is depicted in Fig. 6. It can be noticed that the end-to-end delay increases with increasing speed of nodes. This phenomenon is a consequence of higher mobility causing frequent route changes and frequent link failures. PDTORA maintains the delay QoS within the specified limit (250 ms) and thus has 60% improvement over TORA. End-to-end delay with different mobile speeds for number of nodes from 10 to 50 is shown in Fig. 7.

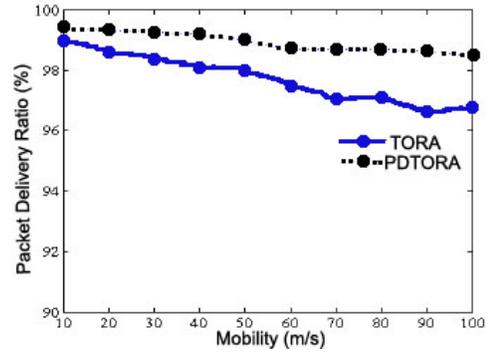

Figure 4. Effect of Mobility on Packet Delivery Ratio

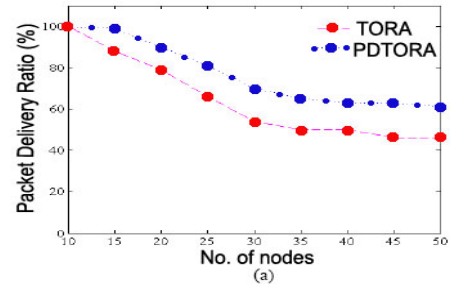

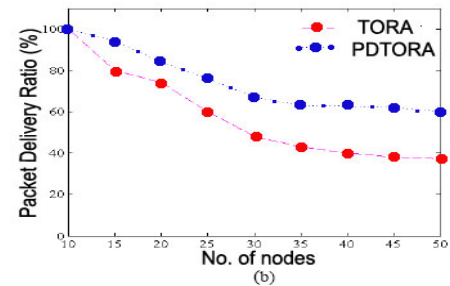

Figure 5. Effect of number of Node on Packet Delivery Ratio (a) mobility 10 m/s ; (b) mobility 20 m/s;

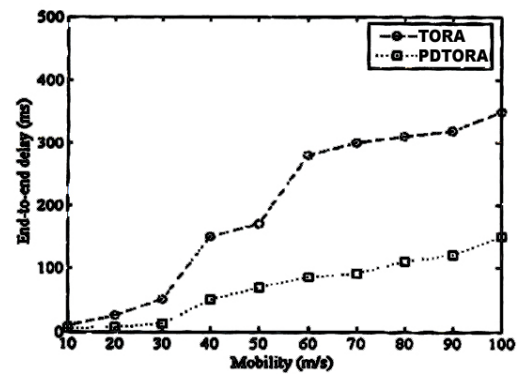

Figure 6. Effect of Mobility on End-To-End Delay for Pause Time of 10s

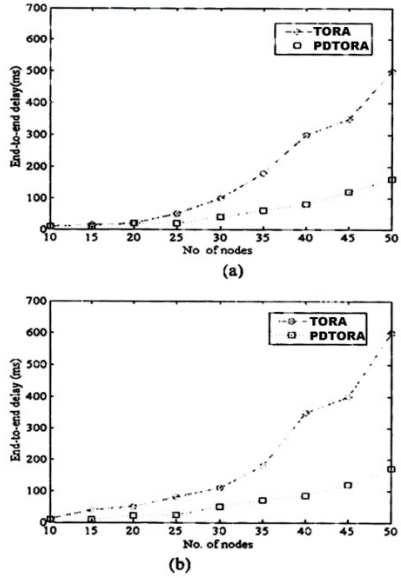

Figure 7. Effect of number of nodes on end-to-end delay  (a) mobility 10 m/s; (b) mobility 20 m/s;

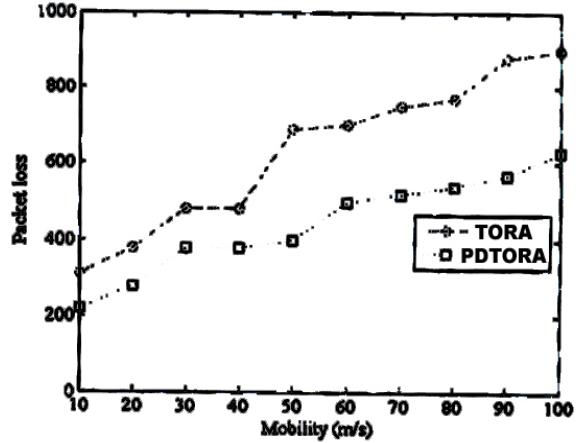

Figure 8. Effect of mobility on packet loss for pause time of 10 s

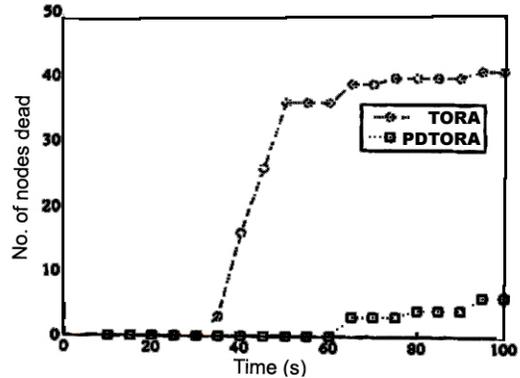

Figure 9. Number of nodes dead vs. Time

It increases with increasing number of nodes, since number of links increase too. In both the cases, PDTORA shows comprehensively better performance over TORA for higher number of mobile nodes.

*4.3. Packet Loss*

Effect of increasing node mobility on packet loss for both TORA and PDTORA is shown in Fig. 8. Packet loss ratio increases with increasing node speed in both the protocols, as a result of more link breakages. However, packet loss ratio in TORA remains much higher than that of PDTORA for the entire scenario. The difference is quite comprehensive at higher mobile speeds.

*4.4. Node Lifetime*

In the course of communication, nodes may happen to die out. Fig. 9 shows the number of nodes which die at some time instants using both TORA and PDTORA. It can be clearly noticed that nodes in TORA die earlier than PDTORA. It happens during forwarding of the query packet, when the power level of an intermediate node is found to be less than that mentioned in the QoS extension for power in the query packet. In TORA, the first node dies at t = 25 sec., whereas in PDTORA, the first node dies at t = 65 sec. Again, at time instant t = 100 sec., 41 nodes die in TORA, whereas only 6 nodes die in PDTORA.

## V. CONCLUSION

The current paper presents an extension of TORA protocol with power and delay aware modification (PDTORA). The nodes in the network which do not satisfy to the QoS requirements of maximum delay and minimum power levels, are eliminated from the route of communication, during query phase. Each intermediate node on receipt of the query packet determines whether to forward it or not, depending on the QoS requirements. At the destination, an update packet is generated. Form the simulations, it could be observed that improvement of QoS metrics in PDTORA over TORA is significant.

## VI. FUTURE WORK

The same approach can be effectively used to improve the QoS metric for other on-demand QoS routing protocols. We are currently working over power and delay aware extensions over Dynamic Source Routing (DSR) protocol.